\documentclass[10pt]{article}
\usepackage[OE]{express}
\usepackage{graphicx}
\usepackage{dcolumn}
\usepackage{amsmath}
\usepackage{amsfonts}
\usepackage{bm}
\usepackage{nicefrac}
\usepackage{color}

\begin{document}
\title{Optimised frequency modulation for continuous-wave optical magnetic resonance sensing using nitrogen-vacancy ensembles}

\author{Haitham A. R. El-Ella\authormark{*}, Sepehr Ahmadi, Adam M. Wojciechowski\authormark{**}, Alexander Huck, and Ulrik L. Andersen}

\address{\authormark{}Department of Physics, Technical University of Denmark, Kongens Lyngby, Denmark. 
\\
\authormark{**} On leave from the Jagiellonian University, Krak\'ow, Poland \\
}

\email{\authormark{*}haitham.el@fysik.dtu.dk} %% email address is required



\begin{abstract}
Magnetometers based on ensembles of nitrogen-vacancy centres are a promising platform for continuously sensing static and low-frequency magnetic fields. Their combination with phase-sensitive (lock-in) detection creates a highly versatile sensor with a sensitivity that is proportional to the derivative of the optical magnetic resonance lock-in spectrum, which is in turn dependant on the lock-in modulation parameters. Here we study the dependence of the lock-in spectral slope on the modulation of the spin-driving microwave field. Given the presence of the intrinsic nitrogen hyperfine spin transitions, we experimentally show that when the ratio between the hyperfine linewidth and their separation is $\gtrsim1/4$, square-wave based frequency modulation generates the steepest slope at modulation depths exceeding the separation of the hyperfine lines, compared to sine-wave based modulation. We formulate a model for calculating lock-in spectra which shows excellent agreement with our experiments, and which shows that an optimum slope is achieved when the linewidth/separation ratio is $\lesssim 1/4$ and the modulation depth is less then the resonance linewidth, irrespective of the modulation function used. 
\end{abstract}


%%%%%%%%%%%%%%%%%%%%%%% References %%%%%%%%%%%%%%%%%%%%%%%%%

%%%%%%%%%%%%%%%%%%%%%%%%%%  body  %%%%%%%%%%%%%%%%%%%%%%%%%%
\section{Introduction}
One of the primary tools for precision measurements in modern experimental physics is phase-sensitive detection. As its principle is straightforward - retrieving a modulated signal component obscured in noise by mixing the total signal with a similarly modulated reference - this detection method results in a versatility that is applicable and implementable in a range of scenarios, from atomic magnetometers \cite{Bell1957}, to single-ion lock-in detection \cite{Kotler2011}, scanning probe microscopy \cite{Salapaka2008}, and optical interferometry and spectroscopy \cite{Mohtar2014}. Phase-sensitive detection is often carried out in conjunction with frequency filtration and AC-amplification, which is collectively termed as lock-in detection \cite{Dicke1946}. For sensing devices, this procedure is usually applied to overcome the problem of increased susceptibility to low-frequency or discrete frequency noise components when attempting to enhance the sensitivity. The main challenge usually lies in identifying suitable system parameters for modulation which lends itself to both maximising the systems response and discriminating it from noise lying outside a given bandwidth. This is pertinent in, for example, low-frequency magnetometry (DC to $\lesssim1$ kHz) where $1/f$ noise and slow drifts stemming from the environment start to dominate \cite{Mateos2015, Clevenson2015}. 

High sensitivity magnetometry has been mainly spurred on by advances in superconductivity and atomic magneto-optics \cite{Robbes2006}, where $<$fT$/\sqrt{\text{Hz}}$ sensitivities have been achieved for a $\sim0.3$ cm$^3$ sampling volume in both pulsed and continuous-wave operating regimes \cite{Kominis2003, Sheng2013}. However, there has been a recent surge of interest in exploring diamond-based magnetometery using ensembles of nitrogen-vacancy (NV) defects, due to the relatively simple technical operation under ambient conditions, their small potential sampling volume (approaching $\mu \text{m}^3$ volumes), and their chemical inertness which allows for direct physical contact with delicate biological systems \cite{Schirhagl2014, Rondin2014}. NV magnetic sensing schemes are usually based on either continuous-wave optically detected magnetic resonance (\textit{cw}-ODMR) or pulsed ODMR, both which are used to determine small magnetic field changes $\delta\textbf{B}$ around a fixed \textbf{B} field offset. \textit{Cw}-ODMR schemes possess field sensitivities limited by the ratio of the spin resonance linewidth to fluorescence contrast and shot-noise level. On the other hand, pulsed techniques such as Ramsey/$\pi$-pulsed schemes result in sensitivities limited by $1/T_2*$, while spin-echo-based schemes possess enhanced field-sensitivities at the expense of limiting their optimised sensitivity to magnetic fields oscillating at frequencies in the order of $1/T_2$ \cite{Taylor2008}.

When sensing low frequency fields (< 1 kHz) or spatial field variations over macroscopic areas, the use of \textit{cw}-ODMR schemes is more technically convenient, which avoids the difficulty facing pulsed schemes in ensuring sufficient uniform intensity and spin control over a macroscopic volume, and involves simply monitoring the spin resonance frequency through shifts in the detected fluorescence level. The efficiency of this scheme is based on how large a change in fluorescence can be generated and how small a change can be detected, for incremental resonance shifts induced by an external magnetic field. The challenge is thus generating the narrowest spectral linewidth during \textit{cw}-driving while ensuring that the fluorescence contrast is as high as possible. This is primarily limited by the number of spins used and the amount of light that is generated and collected, as well as the inherent collective-spin dephasing rates, the optical and microwave (MW) power-related dephasing, the environmental noise fluctuations, and the technical noise in both the detection apparatus and that introduced by power fluctuations occurring at various frequencies in the drive and measurement fields\cite{Wolf2015}. The use of lock-in detection is therefore well suited to overcome many of these issues in order to ensure the intrinsic-sensitivity of the NV system is maintained.

Previous NV-related work which addresses the use of lock-in detection has focused on single NV based sensing through multi-pulsed phase estimation schemes \cite{Nusran2013}, or their incorporation within a scanning-probe based schemes \cite{Schoenfeld2011}. Alternatively, it has also been presented as an integral component for NV ensemble magnetometery using optical fibre-based read-out \cite{Fedotov2014}, using the diamond as a light trapping-waveguide \cite{Clevenson2015}, through the absorption of the spin-singlet state \cite{Jensen2014}, when avoiding the use of a MW drive field \cite{Wickenbrock2016}, or for sensing biologically generated magnetic fields \cite{Barry2016}. However, to our knowledge, there has been no published investigation of the technicalities on optimising the modulation parameters to maximise the slopes in the measured lock-in spectrum. While this technique is conceptually undemanding, the optimum modulation function for NV ensemble magnetometry is not obvious as it is dependent on the systems spin resonance spectrum and its unique response to optical and microwave drive fields. This is especially true as NV systems possesses two/three (depending on the nitrogen isotope) peaks centred around the electron spin resonance frequency, due to the intrinsic hyperfine coupling between the nitrogen nuclear spin and the electron spin, which all respond identically to an external field. 

The work presented in this article therefore investigates NV ensemble-specific ODMR spectra with particular emphasis on delineating an optimum modulation function of the microwave field drive for \textit{cw}-ODMR sensing. This is carried out by first describing the characteristic NV \textit{cw}-ODMR spectrum, followed by a description of frequency modulated lock-in detection. Finally, experimental and theoretical results are compared and discussed, which show that the optimum modulation function and depth are dependent on the linewidth-to-separation ratio of the measured peaks. 

\section{Optically detected magnetic resonance}
\subsection{Continuous-wave spectrum}

\begin{figure}
\centering
\includegraphics[scale = 0.65]{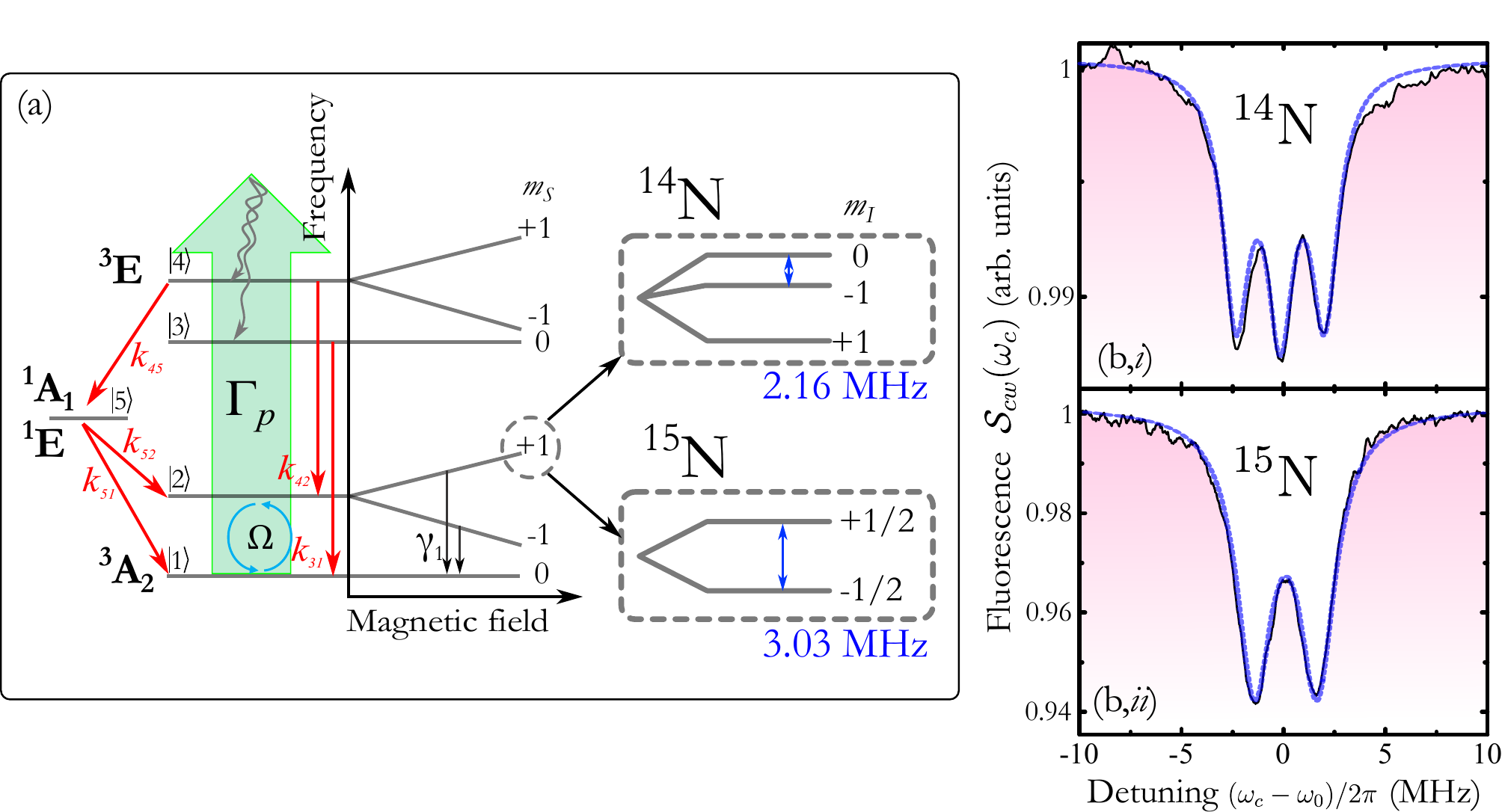}
\caption{(a) A general level schematic of the NV system, used to set up the optical Bloch equations. The electronic structure is comprised of a triplet state ($^{3}$\textbf{A}$_{2}$ $\leftrightarrow$ $ ^{3}$\textbf{E}) and a singlet shelving state ($^{1}$\textbf{E} $\leftrightarrow$ $ ^{1}$\textbf{A}$_1$). The $^{3}$\textbf{A}$_{2}$ state levels can be coherently driven with the application of a MW drive with a Rabi frequency $\Omega$, while above band excitation is carried out with a rate $\Gamma_p$, with the optical excitation processes being almost perfectly spin conserving. The spin levels are split with the introduction of a magnetic field, and couple with the nuclear spin of the N atom, generating an additional two or three hyperfine levels depending on the N isotope. These hyperfine levels are evident in \textit{cw}-ODMR as shown in the measured spectra (b,i,ii) from an NV ensemble, where a weak magnetic field ($\sim5$ mT) has been applied to spectrally separate the distinct crystallographic sub-groups. The dotted blue lines are fits using equation (4) with the parameters $\Gamma_p/2\pi\simeq20$ kHz, $\Omega/2\pi\simeq30$ kHz, and $\gamma_2^*/2\pi\simeq700$ kHz for $^{14}$N, and $\Gamma_p/2\pi\simeq50$ kHz, $\Omega/2\pi\simeq100$ kHz, and $\gamma_2^*/2\pi\simeq 1$ MHz for $^{15}$N.}
\end{figure}

All \textit{cw}-ODMR measurements presented in this article are carried out using an ensemble of native $^{14}$NV$^-$  in an untreated single-crystal diamond (Element 6) grown using chemical vapour deposition (with the exception of the spectrum shown in Fig. 1(b,ii), which was measured from a custom-grown diamond with an isotopically purified nitrogen source to produce an ensemble of $^{15}$NV$^-$). The NV centre is a coupled pair of defects within a diamond carbon lattice, consisting of a substitutional nitrogen atom and a point vacancy. The defect pair exists in both a negatively charged and neutral configuration, as well as in two different isotopic forms of $^{14}$N and $^{15}$N. While both the neutral and charged state fluoresce, it is the negatively charged state which possesses a spin triplet configuration because of its extra electron, resulting in a magnetic-dipole transition with magnetically tunable spin levels \cite{Doherty2013}. The energy levels of the most relevant transitions are shown in Fig. 1(a). Coupled with a near-perfect spin-conserving optical dipole of the NV centre, an optical-contrast spectrum ($\mathcal{S}_{cw}(\omega_c)$, Fig. 1(b)) of the spin resonances can be generated by simultaneously shining light and sweeping the frequency of an applied microwave (MW) field across the spin Larmor frequency. This presents itself as magnetically sensitive probe, which changes its resonance frequency, and therefore its fluorescence rate, as a function of $\gamma_e B_z$, where $\gamma_e \sim28$ MHz/mT is the electron gyromagnetic ratio, and $B_z$ is a magnetic field vector component projected along the crystallographic axis of the NV defect. 

The use of a single or ensemble of spins has direct implications on the apparatus used, and the overall sensitivity. For single NV spins, the amount of light generated is within the $ \sim0.05$ pW range, which falls within the working limits of single photon counting avalanche photodiodes (APD). These usually give excellent signal-to-noise ratios due to their low intrinsic noise levels, and do not require additional lock-in detection provided that the noise in the driving fields or environment do not exceed the shot-noise level. Single spins have a sensitivity ultimately limited to $\sim$ nT/$\sqrt{\text{Hz}}$, when using pulsed sensing protocols (for continuous wave protocols it is in the order of $\sim$ 100 nT/$\sqrt{\text{Hz}}$), and provide nano-scale sensor resolution. If such spatial resolution is not required using an ensemble of spins boosts the sensitivity by $1/\sqrt{N}$, where $N$ is the number of spins used \cite{Taylor2008}. However, dense ensembles ($>1$ ppm NV) prepared through conventional irradiation and annealing techniques usually display a degraded collective ODMR linewidth due to inhomogeneous broadening  and the increased susceptibility to static and driving field inhomogeneities \cite{Acosta2009}. Furthermore, the generated light from an ensemble is usually beyond the working limits of conventional APDs, and the use of \textit{pin}-type photodiodes is necessary which are inherently noisier then APDs. For dense ensembles, fluorescence rates in the mW range can be obtained which usually allows for a shot-noise limited noise floor to be reached, however the degraded linewidth and the introduction of inhomogeneity-related degradation for higher density ensembles \cite{Acosta2009} implies that there may be an optimum balance between NV density and resulting sensitivity. In the low-density case ($\ll1$ ppm NV), an enhanced fluorescence rate can be achieved without too much degradation of the collective coherence/linewidth. However, the inherent \textit{cw}-ODMR contrast for an ensemble of NVs of a single crystallographic orientation is usually around  1-3\% \cite{Rondin2014}, so the working signal-to-noise ratio needs to exceed this limit. Lock-in detection is therefore ideal for tackling these latter issues, usually allowing for a shot-noise limited spectrum to be achieved when optimally driving a low-density ensemble, while maintaining a linewidth that closely resemble those obtained from single NV defects.  

By continuously applying a MW drive on the steepest part of $\mathcal{S}_{cw}(\omega_c)$, and monitoring the level of fluorescence, the presence, frequency, and relative amplitude of a magnetic field can be deduced. The absolute sensitivity $\delta B$ of such a detection process is proportional to $[d_\omega \mathcal{S}_{cw}\gamma_e]^{-1}$, where $d_\omega \mathcal{S}_{cw}$ is the derivative of $\mathcal{S}_{cw}(\omega_c)$ at a particular drive frequency $\omega_c/2\pi$ \cite{Clevenson2015}. The relevant electronic levels that bring about the characteristic $\mathcal{S}_{cw}(\omega_c)$ spectrum are summarised in Fig. 1(a). The system ground state spins are driven by a coherent MW field with a drive frequency  $\omega_c/2\pi$ and a Rabi frequency $\Omega/2\pi$, together with an above-band excitation rate $\Gamma_p/2\pi$, in this case considered for a 532 nm wavelength laser. This wavelength ensures that the NV$^-\rightleftharpoons$ NV$^0$ charge state fluctuation rate is small in comparison to the NV$^-$ fluorescence rate \cite{Aslam2013}. There is an intrinsic hyperfine interaction between the nitrogen nuclear spin and the electron spin of the NV centre which results in two or three hyperfine resonances depending on the nitrogen isotope, as summarised in Fig. 1. The resulting $\mathcal{S}(\omega_c)$ spectra are shown in Fig. 1(b), with the spectra of the sample used throughout the remaining article shown in Fig. 1(b,i).

A schematic of the experimental setup is shown in Fig. 2(a). A signal generator (Stanford Research Systems SG394) is used to deliver a modulated drive frequency $\omega_c/2\pi$ to an antenna placed close to the diamond sample. The diamond is excited with a 532 nm laser (Verdi SLM Coherent), and the fluorescence is collected using a condenser lens, filtered using a long-pass filter with a 600 nm cut-on, and detected using a biased Si detector (Thorlabs DET36A, with a 10 k$\Omega$ load, resulting in a bandwidth of 400 kHz). The excitation volume is in the order of $10^{-2}$ mm$^{3}$ with an estimated number of $10^{9}$ NVs. The measured full-width at half-maximum of an individual hyperfine transition is $\gamma/2\pi\sim1$ MHz, which was well fitted with a linear combination of a Lorentzian and a Gaussian profile (a pseudo-Voigt function) which is $\sim$98\% Lorentzian and $\sim$2\% Gaussian. This indicates that the in-homogeneous broadening is negligible, and the spectra can be confidently analysed using Lorentzian functions. The detector sends the signal to a two-channel digital lock-in amplifier (Stanford Research Systems SR850), while the MW drive frequency $\omega_c/2\pi$ is modulated at $\nu=$ 30 kHz, for which this frequency was confirmed to be in a flat part of the optical power spectrum, and well within the detectors bandwidth. While the lock-in detector has a bandwidth up to 100 kHz, no further gain was achieved in the signal-to-noise ratio when modulating beyond 30 kHz, beyond which also a reduction in the lock-in signal amplitude is observed due to the limitations of the NV ensemble re-polarisation rate (set by the excitation rates used). Simultaneous excitation of all three hyperfine transitions of $^{14}$NV$^-$ is carried out by mixing the modulated MW with a frequency equivalent to the axial hyperfine constant of $^{14}$NV$^-$, which is $A_\parallel=2.16$ MHz. 

The $\mathcal{S}_{cw}(\omega_c)$ spectrum of an NV ensemble can be simulated using a set of optical Bloch equations, by considering a single spin transition (e.g. $\vert m_s= 0\rangle \leftrightarrow\vert$+1$\rangle$) in a total of five levels as shown in Fig. 1(a), with the Hamiltonian: 
\begin{equation}
\hat{\mathcal{H}}/\hbar= \sum^5_i\omega_i\vert i\rangle \langle i\vert - \Omega\cos{(\omega_c t)}\big(\vert 1\rangle \langle 2\vert + \vert 2\rangle \langle 1\vert\big). 
\end{equation}
Only the dominant decay paths considered are highlighted in red in Fig. 1(a), and only the ground-state level transition ($\vert1\rangle  \leftrightarrow\vert2\rangle$) is described in terms of a coherent resonant drive. In a rotating reference frame with $\omega_c$ and using the rotating-wave approximation, the optical Bloch equations are:
\begin{eqnarray}
\dot{\rho}_{11} &&= -\Gamma_{p}\rho_{11}+k_{31}\rho_{33}+k_{41}\rho_{44}+k_{51}\rho_{55}\nonumber
\\\nonumber
&&-\frac{k_{21}}{2}(\rho_{11}-\rho_{22})-\frac{i}{2}\Omega(\rho_{12}-\rho_{21}),
\\
\dot{\rho}_{22} &&= -\Gamma_{p}\rho_{22}+k_{32}\rho_{33}+k_{42}\rho_{44}+k_{52}\rho_{55}\nonumber
\\\nonumber
&&-\frac{k_{21}}{2}(\rho_{22}-\rho_{11})+\frac{i}{2}\Omega(\rho_{12}-\rho_{21}),
\\
\dot{\rho}_{33} &&= \Gamma_{p}\rho_{11}-(k_{35}+k_{32}+k_{31})\rho_{33},
\\\nonumber
\dot{\rho}_{44} &&= \Gamma_{p}\rho_{22}-(k_{45}+k_{42}+k_{41})\rho_{44},
\\\nonumber
\dot{\rho}_{55} &&= k_{45}\rho_{44}+k_{35}\rho_{33} - (k_{52}+k_{51})\rho_{55},
\\\nonumber
\dot{\rho}_{12} &&= -(\gamma_2' - i\delta)\rho_{12} + \frac{i}{2}\Omega(\rho_{22}-\rho_{11}),
\\\nonumber
\dot{\rho}_{21} &&= -(\gamma_2' + i\delta)\rho_{21}     - \frac{i}{2}\Omega(\rho_{22}-\rho_{11}),
\end{eqnarray}
where $\delta=(\omega_c-\omega_0)$ is the detuning between $\omega_c$ and the spin transition frequency $\omega_0$ and $\rho_{ii}$ is the normalised population of a level, while the total dephasing rate is defined as a sum of the longitudinal spin relaxation rate $k_{21}$, pure dephasing rate $\gamma_2^*$, and the optical pump rate through $\gamma_2' = k_{21}/2 + \gamma_2^* +   \Gamma_p/2$. A unit-less, detuning-dependant  \textit{cw-}ODMR fluorescence ratio can then be described in terms of the steady-state populations of the excited states $\vert3\rangle$ and $\vert4\rangle$: 
\begin{eqnarray}
\mathcal{I}_{cw}&=\frac{(k_{31}+k_{32})\rho_{33}^{ss}}{k_{31}+k_{32}+k_{35}}+\frac{(k_{41}+k_{42})\rho_{44}^{ss}}{k_{41}+k_{42}+k_{45}}.
\end{eqnarray}
The exact steady-state solution for equation (3) is given in Appendix A, while the rates used have been extracted from \cite{Robledo2011}. The \textit{cw}-ODMR spectrum including the hyperfine lines is a sum of three individual peaks spaced by the axial hyperfine constant $A_\parallel$ which is 2.16 MHz for $^{14}$N and 3.03 MHz for $^{15}$N \cite{Doherty2013}:
\begin{equation}
\mathcal{S}_{cw}(\omega_c)= R_0\sum_{m_I}\mathcal{I}_{cw}(\delta+m_I 2\pi A_\parallel),
\end{equation}
where $R_0$ is the off-resonance detection rate, and $m_I$ is the nuclear quantum number which spans either $\{-1,0,1\}$ for $^{14}$N or $\{-\tfrac{1}{2},\tfrac{1}{2}\}$ for $^{15}$N. Equation (4) sufficiently reproduces experimentally observed \textit{cw}-ODMR spectra, in particular as it accounts for power-related broadening and the dependence of the linewidth/contrast ratio to the spin excitation and polarisation rates. It is used as a basis for all simulations presented in this article, where the simulated fluorescence contrast and linewidth are an outcome of the given excitation rates, rather than postulated values. Through our measurements, we observe that using this model with rates measured from a single NV \cite{Robledo2011} can successfully represent a low density NV ensemble, and routinely obtains excellent agreement between the generated and measured spectra, as shown in Fig. 1(b,i,ii).

\subsection{Lock-in spectrum}
Generally, an ideally modulated signal can be decomposed as a sum of a static and time-varying component which is a product of a time-independent amplitude and an oscillatory function (i.e. $V(t) =V_0 +  A\cos(2\pi\nu t + \phi(t))$). A modulated \textit{cw}-ODMR signal recorded with a photo-detector can therefore be described to first order as the sum of the unmodulated steady-state spectrum (4) and a product of two complex phasors rotating in opposite directions, with a time-invariant amplitude also defined by (4):
\begin{eqnarray}
\mathcal{S}_{cw}(\omega_c,t)&=\small{\frac{1}{2}}\mathcal{S}_{cw}(\omega_c)+\small{\frac{1}{4}}\mathcal{S}_{cw}(\omega_c )\big[e^{i(2\pi\nu t+\varphi_s)}+e^{-i(2\pi\nu t+\varphi_s)}\big],
\end{eqnarray}
where $\varphi_s$ is the signal phase. Using (5), a lock-in signal $\mathcal{S}_{LI}$ can be described  as a time integrated product of a mixed reference signal $\mathcal{S}_{ref}$ and a measured input signal $\mathcal{S}_{cw}$, both modulated at a frequency  $\nu$, with an amplification gain factor $\mathcal{A}$, as highlighted in Fig. 2(a):
\begin{eqnarray}
\mathcal{S}_{LI}(\omega_c)=\frac{\mathcal{A}\mathcal{S}_{cw}(\omega_c)}{2\tau}\int^{^{\nicefrac{\tau}{2}}}_{_{-\nicefrac{\tau}{2}}}\big(e^{i\phi}+e^{-i(4\pi\nu t+\varphi_s+\varphi_r)}\big)dt,
\end{eqnarray}
where the static component in (5) is removed by the lock-in bandpass filter at its input, and the remaining signal is then re-normalised and multiplied with a sinusoidal reference signal  $\mathcal{S}_{ref}=e^{-i(2\pi\nu t+\varphi_r)}$ with a phase difference $\phi = \varphi_s-\varphi_r$, and integrated in time using a low-pass filter with a time constant $\tau$. The integral results in a complex function of the amplified input signal which can be decomposed into an in-phase ($X$) and quadrature ($Y$) component:
\begin{eqnarray}
X &= \small{\frac{1}{2}}\mathcal{A}\mathcal{S}_{cw}(\omega_c)\cos{(\phi)},
\\
Y &= \small{\frac{1}{2}}\mathcal{A}\mathcal{S}_{cw}(\omega_c)\sin{(\phi)}.
\end{eqnarray}
These can be measured individually using a two-channel lock-in detector, as schematically illustrated in Fig. 2(a). By measuring the quadratures simultaneously  a phase-independent magnitude ($R=\sqrt{X^2+Y^2}$) can also be obtained. The decomposition of (5) and the resulting expression in (7) or (8) takes on different forms depending on the modulation mechanism. In the case of ODMR, modulation can be applied to either the MW drive, the bias magnetic field, or the polarising laser drive. Modulation of the MW drive is a technically convenient approach with respect to versatility and fine control, as it can usually be controlled directly from the MW source without repercussions for the rest of the experimental setup. Frequency modulation (FM) is preferable to amplitude modulation (AM) as it results in a dispersion line-shape of the spin resonance, which possesses its highest sensitivity (point of maximum slope) on resonance where the signal is zero, while also displaying an approximately linear response for small field changes. In contrast, AM results in a Lorentzian/Gaussian line-shape which possesses its highest sensitivity on the line-shapes side where the signal is not zero and therefore more susceptible to fluctuations and noise in the driving fields. An example of measured in-phase $(X)$ lock-in ODMR spectra from an ensemble of $^{14}$NV$^-$ using an AM or FM MW field is shown in Fig. 2(b,i,ii)

\begin{figure}
\centering 
\includegraphics[scale = 0.185]{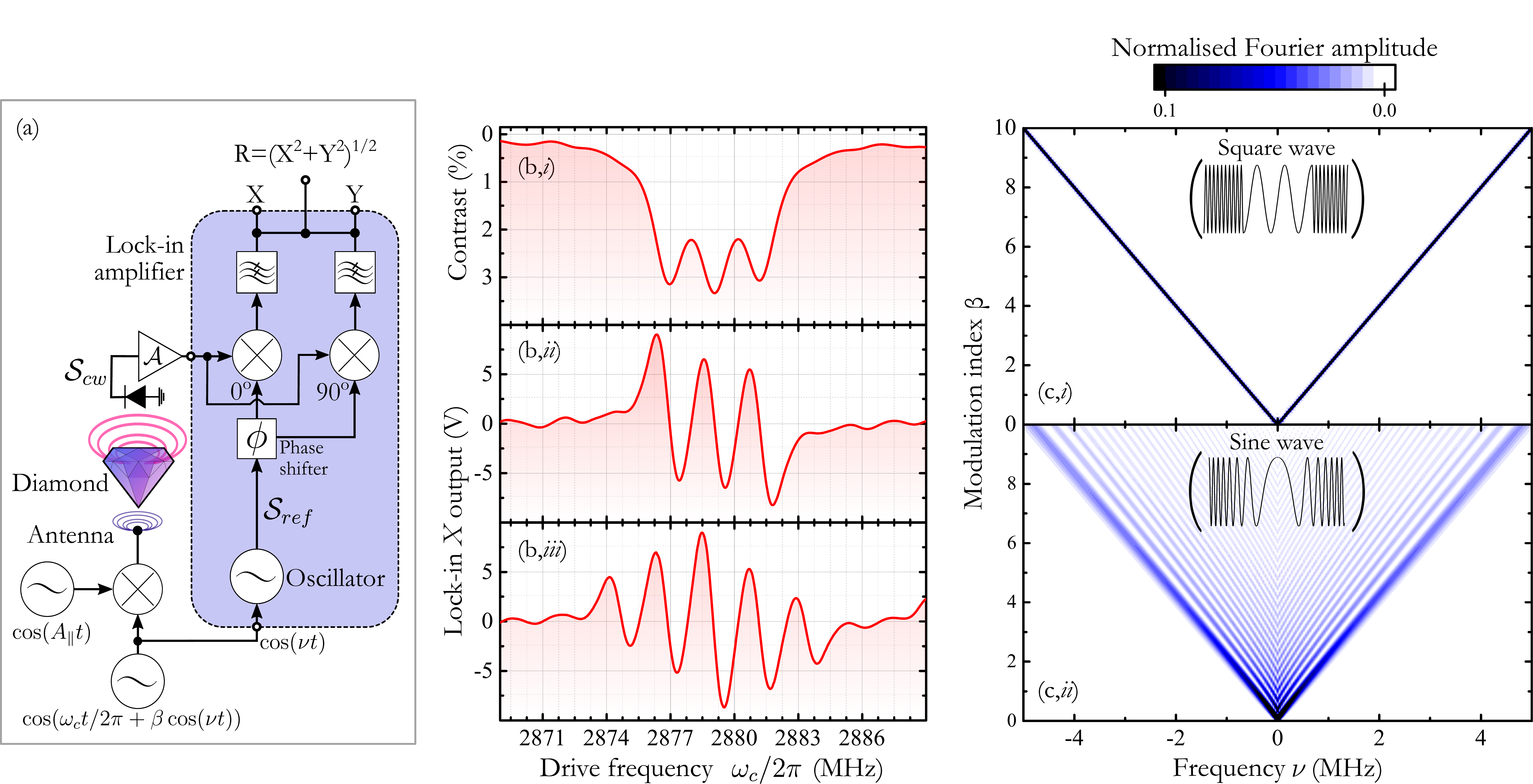}
\caption{(a) Schematic of the experimental setup and the two-channel lock-in detector. (b) Measured spectrum of an ensemble of $^{14}$NV showing (i) in-phase $X$ lock-in spectrum using an amplitude modulated MW field and plotted in terms of the measured fluorescence contrast, (ii) $X$ lock-in spectrum spectrum using sine-wave single-frequency modulation, and (iii) three-frequency excitation of all hyperfine lines. (c) The Fourier spectrum for (i) $B_{\#}(t)$ (10)  and  (ii) $B_{\sim}(t)$ (9) as a function of modulation depth $\Delta\omega$ and index $\beta$, for a fixed modulation frequency $\nu$. The amplitude scale has been reduced to highlight the presence and distribution of peaks in the case of $B_{\sim}(t)$.}
\end{figure}

For FM imposed either through the drive frequency or the external bias magnetic field $B_z$, the detuning $\delta$ becomes a function of time. Such a modulation can be either continuous or discrete, and may be represented by modulation functions that are either sine-wave (continuous, $\sim$) or square-wave (discrete, $\#$), respectively:
\begin{eqnarray}
B_{\sim}(t) &= \cos{(\omega_c t+\beta\sin{(2\pi\nu t)})}=\sum^{+\infty}_{n=-\infty}J_n(\beta)\cos{(\omega_c t+n2\pi\nu t)},
\\
B_{\#}(t)& = \cos\big(\omega_ct+\Delta\omega~\text{sgn}[\cos{(2\pi\nu t)}]t\big),
\end{eqnarray} 
where $J_n$ is a Bessel function of the first kind of order $n$, $\Delta\omega$ is the frequency modulation depth and $ \beta=\Delta\omega/2\pi\nu$ is the modulation index. Analysis of the Fourier spectrum of these two functions, shown in Fig. 2(c),  provides direct insight into their characteristics. The change in the frequency spectrum as function of  $\beta$ for a fixed $\nu$ is stark: $B_{\sim}(t)$ disperses its power over increasingly larger number of frequency components, while for $B_{\#}(t)$ the power remains largely within the two $\Delta\omega$ separated frequencies. Although the bandwidth needed to analytically describe $B_{\sim}(t)$ is infinite, $\sim$99\% of the modulated signal power is present in approximately $n\simeq\lceil\beta\rceil$ frequency components separated by at most $n\nu$ from the central carrier frequency \cite{Carson1922}. With these FM functions in mind, expression (6) can be reformulated in terms of $\mathcal{S}_{cw}(\omega_c(t))$ using a Taylor series expansion about $\omega_c$. This results in an approximation of the in-phase/quadrature components $X/Y$ as the difference between two or $n$ out-of phase spectra separated by $\Delta\omega$ or $\nu$, respectively. When $\phi=0$, the in-phase $X$ output for both modulation functions are:
\begin{eqnarray}
X^{FM}_{\#}(\omega_c)&&\approx\frac{\mathcal{A}}{2}\big(\mathcal{S}_{cw}(\omega_c+\Delta\omega) -\mathcal{S}_{cw}(\omega_c-\Delta\omega)\big),
\\
X^{FM}_{\sim}(\omega_c)&&\approx\frac{\mathcal{A}}{2}\sum_{n=0}^{\lceil\beta/2\rceil}J_n(\beta)\big(\mathcal{S}_{cw}(\omega_c+n2\pi\nu) -\mathcal{S}_{cw}(\omega_c-n2\pi\nu)\big).
\end{eqnarray}
With the presence of the hyperfine transitions and their considerable overlap, an increased contrast can be achieved by simultaneously exciting all of them. The enhancement of the slope is verified to be in the order of 2 - 3 times depending on the excitation and dephasing rates, and is obtained by mixing $\omega_c$  with a frequency equal to the hyperfine separation. For $^{14}$NV, this generates the in-phase $X$ lock-in spectrum shown in Fig. 2(b,\textit{iii}). Analytically, this modifies the expressions (11) and (12) with an additional summation over the number of frequency components $m_x$ (which is equal to $m_I$):   
\begin{eqnarray}
X^{A_\parallel}(\omega_c)&= \sum_{m_x}X^{FM}(\omega_c + m_x 2\pi A_\parallel).
\end{eqnarray}

\section{Modulation function dependence}
\subsection{Single and multiple frequency modulation}

\begin{figure}
\centering 
\includegraphics[scale = 0.5]{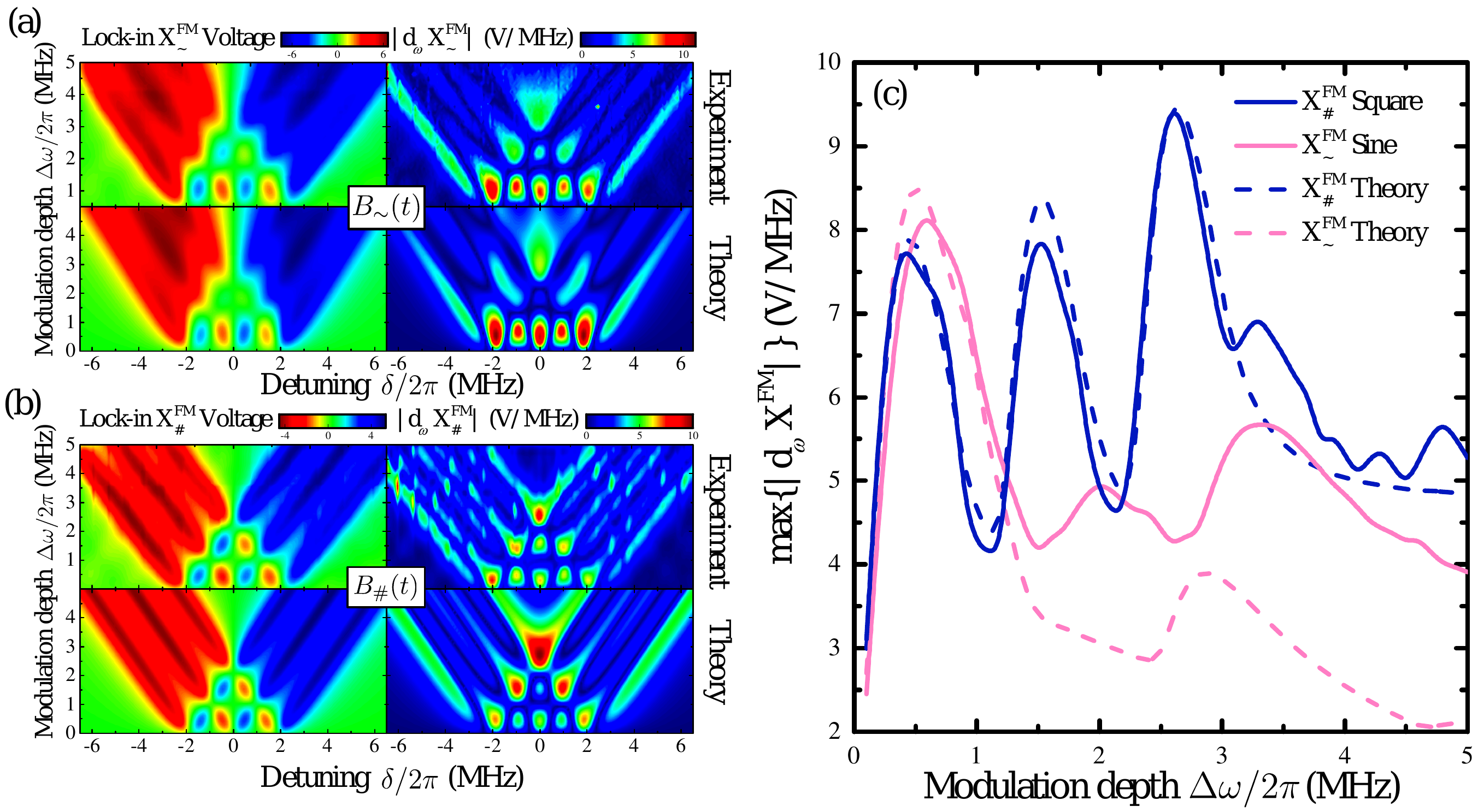}
\caption{Experimental and simulated two-dimensional plots of $X^{FM}(\omega_c,\Delta\omega)$ for (a) sine-wave modulated drive and (b) square-wave modulated drive, shown beside their first-order derivative. (c) Experimental and theoretical values of max$\{\vert d_\omega X^{FM}(\omega_c)\vert\}$ as a function of $\Delta\omega$ for both modulation functions. The simulation parameters used for these Figs. are $\Gamma_p/2\pi\simeq 150$ kHz, $\Omega/2\pi \simeq 300$ kHz, and $\gamma_2^*/2\pi\simeq500$ kHz.}
\end{figure}

Given the finite linewidth $\gamma$ of the individual hyperfine transitions and the constant spectral separation between the hyperfine lines $A_\parallel$, the optimal modulation depth $\Delta\omega$ is expected to occur within the frequency span of all the hyperfine transitions $\sim 2A_\parallel$. Exactly what modulation depth and function is optimal will then depend on $\gamma$, which is dependent on both the intrinsic properties of the diamond, and extrinsically on the excitation rates $\Gamma_p$ and $\Omega$.  

To investigate the dependence of the sine- and square-type modulation functions on the maximum achievable slope, we measure the in-phase $X$ spectrum and its maximum absolute slope max$\{\vert d_\omega X^{FM}\vert\}$ as a function of modulation depth $\Delta\omega$. These are shown in Fig. 3(a) for single-frequency sine, and in Fig. 3(b) for square-wave  modulation, in comparison to simulated spectra using equations (11) and (12). There is very good agreement between the measured and theoretical spectra, particularly as the subtle difference between square-wave and sine-wave modulation is reproducible. Figure 3(c) plots max$\{\vert d_\omega X^{FM}\vert\}$ for either modulation function for both experimental and theoretical trends. These trends show that, for the given ensemble and measurement parameters, sine-wave modulation gives an optimum max$\{\vert d_\omega X^{FM}\vert\}$ when $\Delta\omega/2\pi$ is $\sim0.5\gamma$ of a single hyperfine-transition peak ($\sim1$ MHz), while square-wave modulation exceeds this when $\Delta\omega/2\pi$  is larger than the full span of the three hyperfine lines at around $\Delta\omega/2\pi\sim$2.5 MHz. This difference is due to two factors, which include the ratio between $\gamma$ and the frequency separation ($\xi = \gamma/2\pi A_\parallel$), as well as the spectral characteristics of the two modulation functions. Because for a sine-function the MW power is dispersed across a larger bandwidth as the modulation depth is increased, max$\{\vert d_\omega X^{FM}\vert\}$ is reduced because of the decreasing amplitude. While this should be compensated for by increasing $\nu $ to maintain a constant $\beta$, this is impractical as $\nu$ is usually chosen beforehand on the basis of the inherent noise spectral density and the bandwidth limitations of the apparatus and the system itself. In comparison, square-wave modulation is by definition discrete, and the distribution of power is, to first order, independent of the modulation parameters. A drawback with square-wave modulation is that when mixing with a sine-wave reference, multiple odd harmonics are generated. This degrades the demodulated amplitude by a factor which is dependent on the signal composition, as some of the power is distributed into these odd harmonics. However the power loss through this mechanism is negligible compared to the amplitude decrease when increasing $\Delta\omega$ of a sine-wave modulated spectrum.   

\begin{figure}
\centering 
\includegraphics[scale = 0.5]{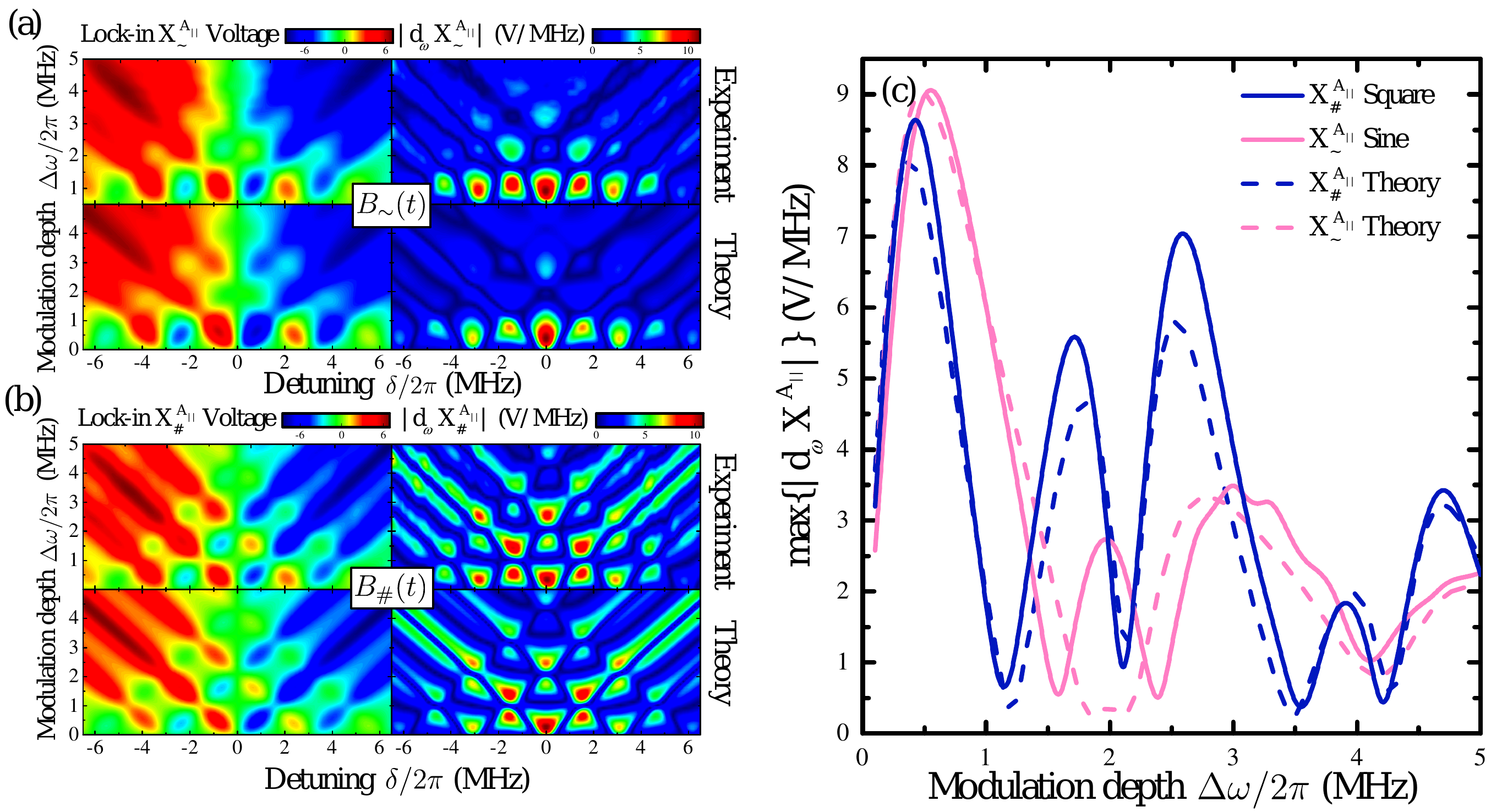}
\caption{Experimental and simulated two-dimensional plots of $X^{A_\parallel}(\omega_c,\Delta\omega)$ for (a) sine-wave modulated drive and (b) square-wave modulated drive, shown beside their first-order derivative. (c) Experimental and theoretical values of max$\{\vert d_\omega X^{A_\parallel}(\omega_c)\vert\}$ as a function of $\Delta\omega$ for both modulation functions. The simulation parameters used for these Figs. are $\Gamma_p/2\pi\simeq 150$ kHz, $\Omega/2\pi \simeq 100$ kHz, and $\gamma_2^*/2\pi\simeq500$ kHz.}
\end{figure}

The in-phase $X$ spectrum generated when all hyperfine lines are simultaneously excited is shown for sine-wave modulation in Fig. 4(a), and for square-wave modulation in Fig. 4(b), in comparison to the simulated spectra using equation (13). In contrast to single-frequency excitation, max$\{\vert d_\omega X^{A_\parallel}\vert\}$ is obtained for a $\Delta\omega$ that is around half of $\gamma$ for both modulation functions and the given ensemble and measurement parameters, as highlighted in Fig. 4(c). The discrepancy in this case is related to the effect of the enhanced contrast on $\vert d_\omega X\vert$ in relation to $\xi$, as discussed in the following section. 

\subsection{Projected modulation function trends}

\begin{figure}
\centering 
\includegraphics[scale = 0.35]{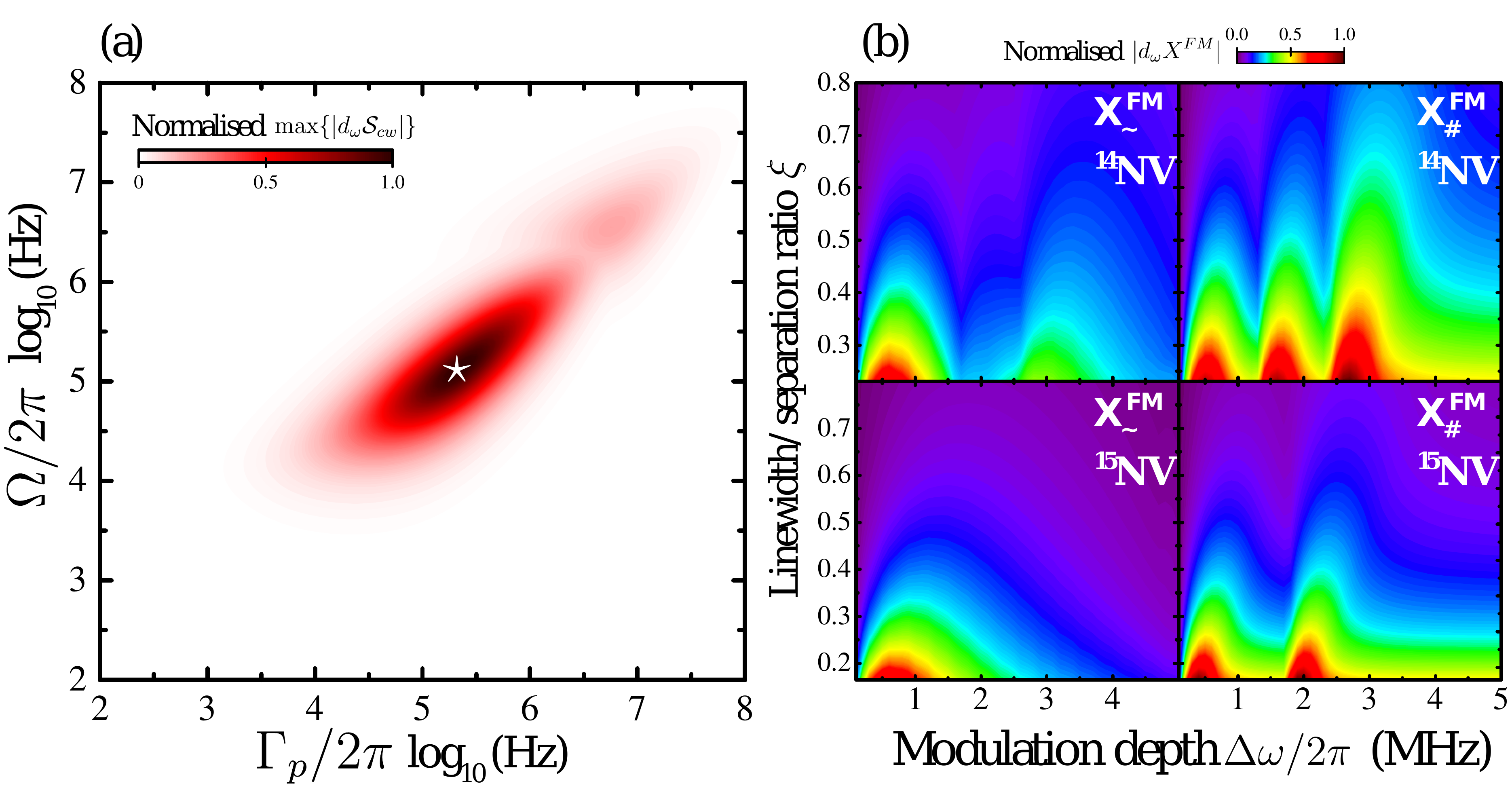}
\caption{(a) Simulation of the steepest slope of expression (4) as a function of Rabi frequency $\Omega$ and optical excitation rate $\Gamma_p$ for $\gamma_2^* /2\pi\simeq 500$ kHz and $k_{21}/2\pi = 1$ kHz. The parameters for the maximum slope, designated by a star, are used for the simulations in (b), which show the normalised maximum slope for single frequency excitation as a function of $\xi$ and modulation depth $\Delta\omega$ for sine ($X_{\sim}^{FM}$) and square ($X_{\#}^{FM}$) wave modulation for both NV isotopes.}
\end{figure}

In order to study how the modulation function influences the lock-in signal slope, the dependence of the unmodulated signal slope on the optical and MW excitation rate needs to be assessed. Prior to lock-in detection, the linewidth $\gamma$ and therefore the slope $\vert d_\omega\mathcal{S}_{cw}\vert$ is dependant on the excitation ratio $\Omega$:$\Gamma_p$, and the effective dephasing rate $\gamma_2'$. In particular max$\{\vert d_\omega \mathcal{S}_{cw}\vert\}$ is obtained when the depletion rate of the spin levels $^{3}$\textbf{A}$_{2}$ by $\Gamma_p$ is large enough to circumvent the effects of MW power broadening, yet below the pure dephasing rate $\gamma_2*$ \cite{Dreau2011, Jensen2013}. This is simulated in Fig. 5(a) which plots $\vert d_\omega\mathcal{S}_{cw}\vert$ as a function of $\Omega$ and $\Gamma_p$, and shows an almost 1:1 correspondence for $\Omega$:$\Gamma_p$ is required to achieve an optimum slope for a given $\gamma^*_2/2\pi\simeq 500$ kHz. The optimum point of max$\{\vert d_\omega \mathcal{S}_{cw}\vert\}$ is highlighted by a star, and using these driving rates, $\vert d_\omega X^{FM}\vert$ is studied as a function of $\xi$ and $\Delta\omega$ for both modulation functions and isotopic forms of the NV center. These are plotted as a function of $\xi$ and $\Delta\omega$ for single-frequency excitation in Fig. 5(b). 

Despite the stark difference between the modulation functions excitation spectrum (shown in Fig. 2(c)), it is the ratio $\xi$ which is observed to be the limiting factor when generating the maximum obtainable slope max$\{\vert d_\omega X^{FM}\vert\}$. As $\xi$ approaches 1/4 for both $^{14}$NV and $^{15}$NV, the inherent overlap of the hyperfine lines degrades the $\vert d_\omega X^{FM}\vert$ values in the region between the peaks, while increasing the overall amplitude and therefore enhances $\vert d_\omega X^{FM}\vert$ on the sides of the peak generated by the sum of all the hyperfine lines. As such, for $\xi>1/4$, square-wave modulation outperforms sine-wave modulation, and the optimum $\Delta\omega$ will be the full span of all the hyperfine transitions. For $\xi<1/4$, the difference between square- and sine-wave modulation is negligible, and the optimum $\Delta\omega$ remains within $\gamma$ for both modulation functions, at values equivalent to $A_\parallel$ for square-wave modulation. Given the inherent $A_\parallel$ values of NVs, these ratios translate into $\gamma/2\pi$ of $\sim0.5$ MHz for $^{14}$NV, and  $\sim0.8$ MHz for $^{15}$NV. Ultimately, this implies that square-wave modulation is the optimum choice when $\gamma$ exceeds these values, to maximise $\vert d_\omega X(\omega_c)\vert$, and thereby the achievable sensitivity $\delta B \propto [d_\omega X(\omega_c)\gamma_e]^{-1}$.

\begin{figure}
\centering
\includegraphics[scale = 0.375]{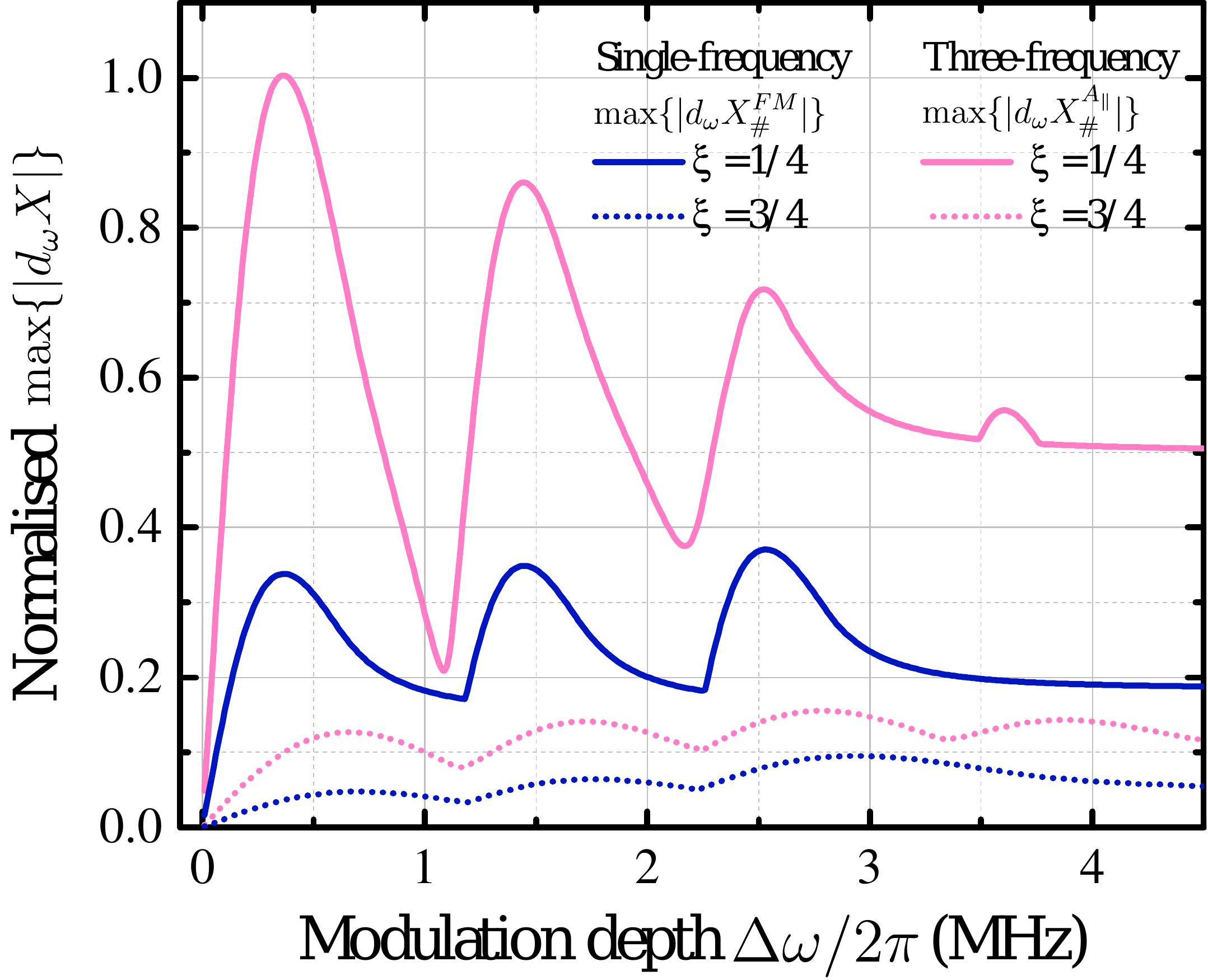}
\caption{Comparison of simulated max$\{\vert d_\omega X_\#\vert\}$ for single- and three-frequency excitation of an ensemble of $^{14}$NV, using two different $\xi$ values. The normalisation is with respect to the maximum simulated slope, and the $\xi$ values correspond to a linewidth of $\sim0.5$ MHz and $\sim1.6$ MHz.}
\end{figure}

From these simulations, it is evident that the optimum $\Delta\omega$ is dependent on the relative variation of $\vert d_\omega X^{FM}\vert$ within the spectrum: the more 'smeared' a spectrum is, the larger $\Delta\omega$ is needed to optimise  $\vert d_\omega X^{FM}\vert$. As such, exciting all three transitions increases the contrast without significantly affecting the relative $\vert d_\omega X^{FM}\vert$ throughout the spectrum, thereby offsetting $\xi$ to larger values. For simultaneous hyperfine level excitation, a larger $\xi$ ratio is therefore required for square-wave modulation to be advantageous. The simulations highlight a threshold that is increased by a factor proportional to the number of excited hyperfine lines $m_I$. This behaviour is highlighted in Fig. 6 where max$\{\vert d_\omega X^{A_\parallel}_{\#}\vert\}$ occurs at a modulation depth that is larger than the inherent linewidth only when $\xi >3/4$. Ultimately, it should be emphasised that with an increase in $\xi$, max$\{\vert d_\omega X\vert\}$ is inherently degraded. The optimum scenario will therefore always occur with three-frequency excitation and ensembles possessing as low a $\xi$ ratio as possible, for which there is no advantage in using either modulation function.

\section{Conclusion}
In this work we identify the optimum lock-in modulation parameters for obtaining the steepest spectral slope in the ODMR spectra measured from a low-density ensemble of $^{14}$NV, for the purpose of \textit{cw}-sensing of magnetic fields. Experimental measurements of the in-phase lock-in spectra from a $^{14}$NV ensemble as a function of modulation depth and modulation function highlighted the advantage of square-wave modulation for the given sample. This also provided an assessment check for the spectral simulations based on a five-level set of optical Bloch equations and its reformulation in terms of a modulated drive frequency. Through accounting for the spectral difference of the modulation functions, this model was able to accurately reproduce the experimentally observed subtleties in the lock-in spectra. Their correspondence provides confidence to the simulated projections, which highlight the key relationship between $\Delta\omega$ and $\xi$. In particular, this showed that for single-frequency excitation when $\xi\gtrsim1/4$, square-wave modulation is optimal with a $\Delta\omega$ which spans the sum of all the hyperfine linewidths. Below this threshold, there is no significant advantage in using either modulation functions. For NVs, this translated into linewidth limits of $\sim0.5$ MHz for $^{14}$NV, and $\sim0.8$ MHz for $^{15}$NV. Ultimately, the model indicates that an optimum slope will always be achieved with as low a $\xi$ ratio as possible, with multi-frequency excitation and a modulation depth that is within the hyperfine linewidth, irrespective of the modulation function. 

It is anticipated this analysis will benefit the development and optimisation of NV-based sensing schemes and devices, while the simplicity of the presented model should allow for its easy re-modification for alternative systems with different spectral properties. This is envisaged especially as many off-the-shelf NV ensembles display poor $\xi$ ratios, and the production of optimised isotopically-pure diamonds is resource and time intensive. Aside from this, the relationship delineated here may be relevant to any spectral measurement or \textit{cw}-sensing scheme which uses closely-spaced spectral features that respond identically to external perturbations. 

\appendix
\section{Analytical solution to steady-state fluorescence ratio derived from the 5-level Bloch equations}
The steady-state solutions for equation (3) detailed in section two of the manuscript (using notation with respect to Fig. 1(a)) is given as:
\begin{eqnarray}
\mathcal{I}_{cw}&=\frac{\Gamma_p(k_{31}+k_{32})}{K_3^2} \Bigg[1+\Xi+ \frac{\Gamma_p}{K_3}+\frac{\Gamma_p\Xi}{K_4}+\frac{k_{35}\Gamma_p}{K_3K_5}+\frac{k_{45}\Gamma_p\Xi}{K_5 K_4}\Bigg]^{-1}\nonumber
\\
&~~~~~~~~~~+\frac{\Gamma_p(k_{41}+k_{42})}{K_4^2}\Bigg[1+\frac{1}{\Xi}+ \frac{\Gamma_p}{K_4}+\frac{\Gamma_p}{K_3\Xi}+\frac{k_{45}\Gamma_p}{K_4K_5}+\frac{k_{35}\Gamma_p}{K_5 K_3\Xi}\Bigg]^{-1},
\end{eqnarray}
where
\begin{eqnarray}
\Xi &= \frac{\Bigg[   \big(\frac{k_{21}}{2}\big)    +\bigg(\frac{\Gamma_p(k_{32}K_5+k_{52}k_{35})}{K_3 K_5}\bigg)  +\bigg(\frac{\Omega^2\gamma_2'}{2(\gamma_2'^2+\Delta^2)}\bigg)\Bigg] }{ \Bigg[\Gamma_p+\big(\frac{k_{21}}{2}\big)  - \bigg(\frac{\Gamma_p(k_{42}K_5+k_{52}k_{45})}{K_4 K_5}\bigg) +      \bigg(\frac{\Omega^2\gamma_2'}{2(\gamma_2'^2+\Delta^2)}\bigg)\Bigg]},
\\\nonumber
\\
K_3 &=k_{31}+k_{32}+k_{35},~~~~K_4 =k_{41}+k_{42}+k_{45},~~~~K_5 =k_{51}+k_{52}´.
\end{eqnarray} 

\section*{Funding}
Danish Innovation Foundation (EXMAD, the Qubiz centre); Danish Research Council (Sapere Aude DIMS).

\section*{Acknowledgements}
We would like to thank Kristian Hagsted Rasmussen for help with diamond sample preparation. 

\begin{thebibliography}{99}

\bibitem{Bell1957} W. E. Bell and A. L. Bloom ``Optical detection of magnetic resonance in alkali metal vapour,'' Phys. Rev. {\bfseries 107}, 1559-1565 (1957).

\bibitem{Kotler2011} S. Kotler, N. Akerman, Y. Glickman, A. Keselman, and R. Ozeri, ``Single-ion quantum lock-in amplifier,'' \nat {\bfseries 473}, 61-65 (2011).

\bibitem{Salapaka2008} S. M. Salapaka and M. V. Salapaka ``Scanning probe microscopy,'' IEEE Control Syst. Mag. {\bfseries 28}, 65-83 (2008)

\bibitem{Mohtar2014} A. Al Mohtar, J. Vaillant, Z. Sedaghat, M. Kazan, L. Joly, C. Stoeffler, J. Cousin, A. Khoury, and A. Bruyant ``Generalised lock-in detection for interferometry: application to phase sensitive spectroscopy and near-field nanoscopy,'' Opt. Express {\bfseries 22}, 22232-22245 (2014)

\bibitem{Dicke1946} R. H. Dicke ``The measurement of thermal radiation at microwave frequencies,'' Rev. Sci. Instrum. {\bfseries 17}, 268-275 (1946).

\bibitem{Mateos2015} I. Mateos, B. Patton, E. Zhivun, D. Budker, D. Wurm, and J. Ramos-Castro ``Noise characterization of an atomic magnetometer at sub-millihertz frequencies,'' Sensor Actuat. A-Phys. {\bfseries 224}, 147-155 (2015).

\bibitem{Clevenson2015} H. Clevenson, M. E. Trusheim, T. Schroder, C. Teale, D. Braje, and D. Englund ``Broadband magnetometry and temperature sensing with a light trapping diamond waveguide,'' Nat. Phys. {\bfseries 11}, 393-397 (2015).

\bibitem{Robbes2006} D. Robbes ``High sensitive magnetometers-a review,'' Sensor Actuat. A-Phys. {\bfseries 129}, 86-93 (2006).

\bibitem{Sheng2013} D. Sheng, S. Li, N. Dural, and M. V. Romalis ``Subfemtotesla scalar atomic magnetometry using multipass cells,'' Phys. Rev. Lett. {\bfseries 110}, 160802 (2013).

\bibitem{Kominis2003} I. K. Kominis, T. W. Kornak, J. C. Allred, and M. V. Romalis ``A subfemtotesla multichannel atomic magnetometer,'' Nature {\bfseries 442}, 596-599 (2003).

\bibitem{Rondin2014} L. Rondin, J.-P. Tetienne, T. Hingant, J.-F. Roch, P. Maletinsky, and V. Jacques  ``Magnetometry with nitrogen-vacancy defects in diamond,'' Rep. Prog. Phys. {\bfseries 77}, 056503 (2014).

\bibitem{Schirhagl2014} R. Schirhagl, K. Chang, M. Loretz, and C. L. Degen ``Nitrogen-vacancy centres in diamond: nanoscale sensors for physics and biology,'' Annu. Rev. Phys. Chem. {\bfseries 65}, 83-105 (2014).

\bibitem{Taylor2008} J. M. Taylor, P. Cappellaro, L. Childress, L. Jiang, D. Budker, P. R. Hemer, A Yacoby, R. Walsworth, and M. D. Lukin  ``High-sensitivity diamond magnetometer with nanoscale resolution,'' Nat. Phys. {\bfseries 4}, 810-816 (2008).

\bibitem{Wolf2015} T. Wolf, P. Neumann, K. Nakamura, H. Sumiya, T. Ohshima, J. Isoya, and J. Wrachtrup ``Subpicotesla diamond magnetometry,'' Phys. Rev. X {\bfseries 5}, 041001 (2015).

\bibitem{Nusran2013} N.M. Nusran and M. V. Gurudev Dutt ``Dual-channel lock-in magnetometer with a single spin diamond,'' Phys. Rev. B {\bfseries 88}, 220410(R) (2013).

\bibitem{Schoenfeld2011} R. S. Schoenfeld and W. Harneit ``Real time magnetic field sensing and imaging using a single spin in diamond,'' Phys. Rev. Lett. {\bfseries 106}, 030802 (2011).

\bibitem{Fedotov2014} I. V. Fedotov, L. V. Doronina-Amitonoa, A. A. Voronin, A. O. Levchenko, S. A. Zibrov, D. A. Sidorov-Biryukov, A. B. Fedotov, V. L. Velichansky and A.M. Zheltikov ``Electron spin manipulation and readout through an optical fibre'' Sci. Rep. {\bfseries 4}, 5362 (2014).

\bibitem{Jensen2014} K. Jensen, N. Leefer, A. Jarmola, Y. Dumiege, V. M. Acosta, P. Kehayias, B. Patton, and D. Budker ``Cavity-enhanced room-temperature magnetometry using absorption by nitrogen-vacancy centres in diamond,'' Phys. Rev. Lett. {\bfseries 112}, 160802 (2014).

 \bibitem{Wickenbrock2016} A. Wickenbrock, H. Zheng, L. Bougas, N. Leefer, S. Afach, A. Jarmola, V. M. Acosta, and D. Budker ``Microwave-free magnetometry with nitrogen-vacancy centres in diamond,'' Appl. Phys. Lett. {\bfseries 109}, 053505 (2016).

\bibitem{Barry2016} J. F. Barry, M. J. Turner, J. M. Schloss, D. R. Glenn, Y. Song, M. D. Lukin, H. Park, and R. L. Walsworth ``Optical magnetic detection of single-neuron action potentials using quantum defects in diamond,'' Proc. Natl. Acad. Sci. USA  {\bfseries 113}, 14133-14138 (2016).

\bibitem{Doherty2013}  M. W. Doherty, N. B. Manson, P. Delaney, F. Jelezko, J. Wrachtrup and L. C. L. Hollenberg ``The nitrogen-vacancy color centre in diamond,'' Phys. Rep. {\bfseries 528}(1), 1-45 (2013).

\bibitem{Acosta2009} V. M. Acosta, E. Bauch, M. P. Ledbetter, C. Santori, K.-M. C. Fu, P. E. Barclay, R. G. Beausoleil, H. Linget, J. F. Roch, F. Treussart, S. Chemerisov, W. Gawlik, and D. Budker ``Diamonds with a high density of nitrogen-vacancy centres for magnetometry applications,'' Phys. Rev. B {\bfseries 80}, 115202 (2009).  

\bibitem{Aslam2013} N. Alsam, G. Waldherr, P. Neumann, F. Jelezko, and J. Wrachtrup ``Photo-induced ionization dynamics of the nitrogen vacancy defect in diamond investigated by single-shot charge state detection,'' New Journal of Physics  {\bfseries 15}, 013064 (2013).

\bibitem{Robledo2011}  L. Robledo, H. Bernien, T. V. D. Sar, and R. Hanson ``Spin dynamics in the optical cycle of single nitrogen-vacancy centres in diamond,'' New Journal of Physics {\bfseries 13}, 025013 (2011).

\bibitem{Carson1922} J. R. Carson ``Notes on the theory of modulation,'' Proceedings of the Institute of Radio Engineers {\bfseries 10}, 57-64 (1922).

\bibitem{Dreau2011} A. Dr\'{e}au, M. lesik, L. Rondin, P. Spinicelli, O. Arcizet, J.-F- Roch, and V. Jacques ``Avoiding power broadening in optically detected magnetic resonance of single NV defects for enhanced dc magnetic field sensitivity,'' \prb {\bfseries 84}, 195204 (2011).

\bibitem{Jensen2013}  K. Jensen, V. M. Acosta, A. Jarmola, and D. Budker, ``Light narrowing of magnetic resonances in ensembles of nitrogen-vacancy centres in diamond,'' \prb {\bfseries 87}, 014115 (2013).
\end{thebibliography}
\end{document}